\documentclass[twocolumn,floatfix, showpacs]{revtex4}

\usepackage[final]{epsfig}
\usepackage{amsmath}
\usepackage{parskip}

\begin{document}

\title{Prospects of medium tomography using hard processes inside a soft medium}

\author{Thorsten Renk}
\email{trenk@phys.jyu.fi}
\affiliation{Department of Physics, PO Box 35 FIN-40014 University of Jyv\"askyl\"a, Finland}
\affiliation{Helsinki Institut of Physics, PO Box 64 FIN-00014, University of Helsinki, Finland}

\pacs{25.75.-q,25.75.Gz}

\begin{abstract}
Hard processes leading to high transverse momentum hadron production are calculable in perturbative Quantum Chromodynamics (pQCD) for proton-proton collisions. In heavy-ion collisions, such processes occur as well, and due to a separation of scales are calculable on the level of the production of high $p_T$ partons. These subsequently interact with the medium which originates from  semi-hard and soft processes in the same collision before they hadronize. The presence of the medium thus modifies the momentum spectrum of outgoing partons and thus hard hadrons. The idea of jet tomography is to study this modification in order to learn about the medium density distribution. While this is a compelling idea, its practical application encounters some difficulties. In this paper, the capability of various observables to provide tomographic information about the medium is discussed and compared.
\end{abstract}

\maketitle

\section{Introduction}
\label{sec_introduction}

The expression 'jet tomography' often used to describe the analysis of hard pQCD processes taking place inside the soft matter created in an ultrarelativistic heavy-ion collision. In particular the focus is on the nuclear suppression of hard hadrons in A-A collisions compared with the scaled expectation from p-p collisions due to loss of energy from the hard parton by interactions with the soft medium (see e.g. \cite{Tomo1,Tomo2,Tomo3,Tomo4,Tomo5}). As the use of the word 'tomography' suggests, the ultimate hope here is to make a measurement of the density profile of the soft medium.

The main underlying idea of jet tomography is to start with a hard process which is believed to be well understood and calculable in vacuum. Such a process may be the production of high transverse momentum ($p_T$) hadrons $h$, which in p-p collisions can be obtained in leading order perturbative QCD (LO pQCD) from the (schematical) convolution

\begin{equation}
\begin{split}
d\sigma^{NN \rightarrow h+X} =
&\sum_{fijk}  f_{i/N}(x_1,Q^2) \otimes  f_{j/N}(x_2, Q^2) \\ & \otimes \hat{\sigma}_{ij \rightarrow f+k}  \otimes  D_{f\rightarrow h}^{vac}(z, \mu_f^2)
\end{split}
\end{equation}

where $f_{i/N}(x,Q^2)$ are the non-perturbative distribution functions of a parton $i$ to carry a momentum fraction $x$ of the nucleon $N$ when the distribution is resolved at a scale $Q^2$ (see e.g. \cite{CTEQ1,CTEQ2}), $\hat{\sigma}_{ij \rightarrow f+k}$ are the elementary cross sections for partonic scattering $ij \rightarrow fk$ calculable in pQCD and $D_{f\rightarrow h}^{vac}(z, \mu_f^2)$ is the non-perturbative fragmentation function describing the hadronization of a parton $f$ into a hadron $h$ which carries the momentum fraction $z$ of the original parton when the hadronization scale is $\mu^2$ (see \cite{KKP,AKK}).

If this process is placed into a medium, in principle all of these ingredients may be modified. It has to be remembered that the medium is created at the same time as the hard parton pair by semi-hard and soft inelastic scattering processes between the nucleons of the colliding ions. However, if the momentum transfer in the hard partonic scattering is sufficiently high, then the separation in scale between hard scattering vertex and typical soft medium scale can be used to argue that $\hat{\sigma}_{ij \rightarrow f+k}$ should not be influenced by the presence of a medium as the spatial resolution scale of the hard process is too small to detect medium constituents. 

On the other hand, it is well known that the parton distribution inside a nucleon being part of a nucleus is distinct from $f_{i/N}(x,Q^2)$ discussed above \cite{NPDF,EKS98,EKS07}. However, the change of the distribution can be inferred from e-A or p-A scattering experiments and can hence be assumed to be sufficiently known in heavy-ion collisions.

The medium is then first relevant for the parton emerging from the hard vertex. In principle, one can assume that the presence of the medium modifies the hadronization process of the parton and introduce a medium-modification of the fragmentation function (see e.g. \cite{XN1,XN2}). It is experimentally well known that hadronization processes other than fragmentation, possibly coalescence \cite{Coalescence} or parton recombination \cite{Reco} dominate hadron production below 6 GeV transverse momentum. However, when the hadron is created at sufficiently high $p_T$ it can be assumed that hadronization takes place outside the medium. A rough estimate for the formation distance of a hadron is given by $p_T/m_{had}^2$ which for a 10 GeV pion evaluates to $O(100)$ fm, i.e. the assumption that the hadronization is not influenced by the presence of a medium is rather safe.

Under these conditions, the medium effect on the process can be cast into the form of a probability density $P(\Delta E)$ of the parton experiencing the energy loss $\Delta E$ while traversing the medium. This distribution can be dependent on the parton color charge (i.e. distinct for quarks and gluons) and can in principle also depend on the initial parton energy. 

Thus, in the presence of a medium, for sufficiently hard processes the production of high $p_T$ hadrons can be written as 

\begin{equation}
\label{E-med}
d\sigma_{med}^{AA\rightarrow h+X} \negthickspace = \sum_f d\sigma_{vac}^{AA \rightarrow f +X} \otimes P_f(\Delta E, E) \otimes
D_{f \rightarrow h}^{vac}(z, \mu_F^2)
\end{equation}

where

\begin{equation}
\label{E-2Partons}
d\sigma_{vac}^{AA \rightarrow f +X} = \sum_{ijk} f_{i/A}(x_1,Q^2) \otimes f_{j/A}(x_2, Q^2) \otimes \hat{\sigma}_{ij 
\rightarrow f+k}.
\end{equation}

In these expressions, $f_{i/A}(x_1,Q^2)$ now denote the nuclear parton distribution functions \cite{NPDF,EKS98} and $P_f(\Delta E, E)$ is the probability of energy loss to the medium. In this kinematic limit, the tomographic information on the medium is therefore contained in $P_f(\Delta E, E)$. A measurement of high $p_T$ hadron production can provide information to constrain this quantity. However, in order to obtain tomographic information of the medium, it has to be realized that $P_f(\Delta E, E)$ depends on both the precise nature of the interaction of parton and medium and on the density distribution of the medium. Thus, in order to obtain medium density information, the energy transfer from parton to medium must be assumed to be known.

In the following, we will elaborate on the concepts outlined above. First, we will demonstrate that the energy loss probability is not well constrained by the suppression seen in single hadron spectra. We will then turn to more differential measurements and show what additional information can be obtained to better constrain the energy loss probability distribution and what can be concluded with regard to the density distribution if a specific model for energy loss is assumed. 

\section{The nuclear suppression factor}

The most commonly studied observable in the context of parton energy loss is the nuclear suppression factor

\begin{equation}
\label{E-R_AA}
R_{AA}(p_T,y) = \frac{d^2N^{AA}/dp_Tdy}{T_{AA}(0) d^2 \sigma^{NN}/dp_Tdy}.
\end{equation}

This quantity is unity if an A-A collision can be viewed as a superposition of independent p-p collisions (i.e. no medium is created). Any deviation from unity indicates a non-trivial effect which has to be attributed to either energy loss or the nuclear parton distributions. It has been shown that the effect of the latter is small \cite{EH}. On the other hand, experimentally $R_{AA}$ for identified pions as a function of $p_T$ seems to be flat out to 20 GeV and has a numerical value of about 0.2, i.e. the suppression of high $p_T$ hadrons is substantial \cite{PHENIX_R_AA}.

Since $R_{AA}$ does not contain any spatial information, the production vertices of hard partons and their path through the medium have to be averaged out. Hard vertices $(x_0,y_0)$ are distributed according to a probability density

\begin{equation}
\label{E-PGeo}
P(x_0,y_0) = \frac{T_{A}({\bf r_0 + b/2}) T_A(\bf r_0 - b/2)}{T_{AA}({\bf b})},
\end{equation}

where ${\bf b}$ is the impact parameter. The thickness function is given by the nuclear density
$\rho_A({\bf r},z)$ as $T_A({\bf r})=\int dz \rho_A({\bf r},z)$. Hence, if we can compute the energy loss probability distribution $P_f(\Delta E, E)_{path}$ for a given path through the medium, the quantity to be folded with the pQCD parton spectrum and the fragmentation function is the average over all possible points of origin and paths, i.e.

\begin{equation}
\begin{split}
\label{E-Pav}
\langle P_f(\Delta E, E)\rangle_{T_{AA}} \negthickspace = &\\ \negthickspace \frac{1}{2\pi} \int_0^{2\pi} 
\negthickspace \negthickspace \negthickspace d\phi 
\int_{-\infty}^{\infty} &\negthickspace \negthickspace \negthickspace \negthickspace dx_0 
\int_{-\infty}^{\infty} \negthickspace \negthickspace \negthickspace \negthickspace dy_0 P(x_0,y_0)  
P_f(\Delta E, E)_{path}
\end{split}
\end{equation}

Any information about the medium density is now solely contained in $P_f(\Delta E, E)_{path}$.

\section{Sensitivity of $R_{AA}$}

Before we turn to a specific model of the interaction of the hard parton with the medium, let us discuss how well the functional form of $\langle P_f(\Delta E, E)\rangle_{T_{AA}}$ is constrained by current data. This can be done by inserting a trial ansatz for $\langle P_f(\Delta E, E)\rangle_{T_{AA}}$  into Eq.~(\ref{E-med}) \cite{Gamma-Hadron}.

In general, for a parton with given energy $E$, the effect of $\langle P_f(\Delta E, E)\rangle_{T_{AA}}$ may lead to the following three possibilities: 1) the parton escapes without energy loss (transmission) 2) the parton experiences so much energy loss that it is shifted to a thermal range, i.e. it can no longer be considered hard but becomes part of the bulk medium (absorption) and 3) the parton momentum is shifted, but it still remains sufficiently hard after energy loss (shift). The definition of what $\Delta E$ leads to shift and what is absorption naturally depends on the initial parton energy $E$. This is an implicit dependence of the energy loss probability which is relevant even if $P_f(\Delta E)_{path}$ is itself independent of $E$. In addition, there may also be an explicit dependence of the functional form of the energy loss probability on the initial parton energy.

\begin{figure*}[htb]
\epsfig{file=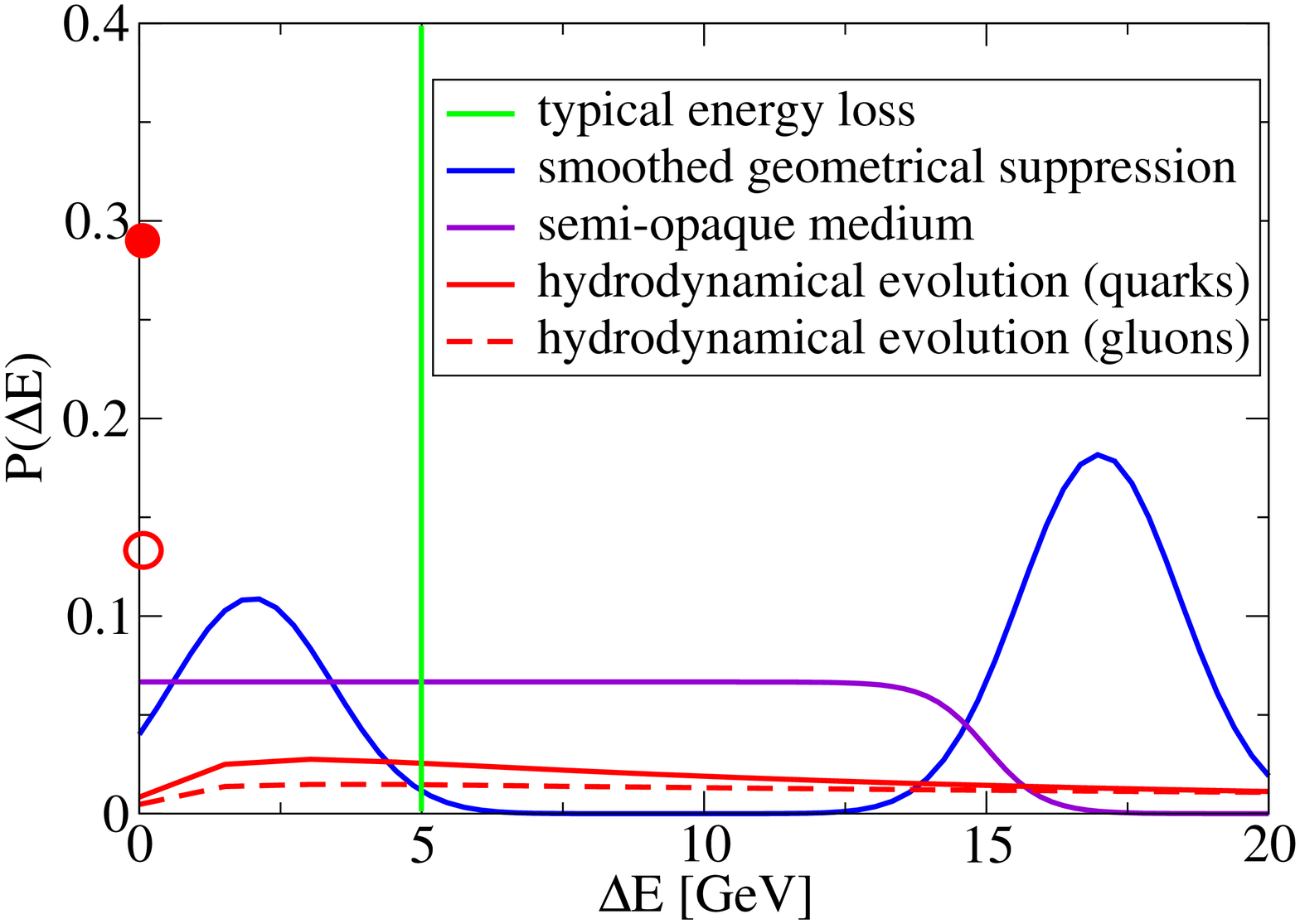, width=8cm} \epsfig{file=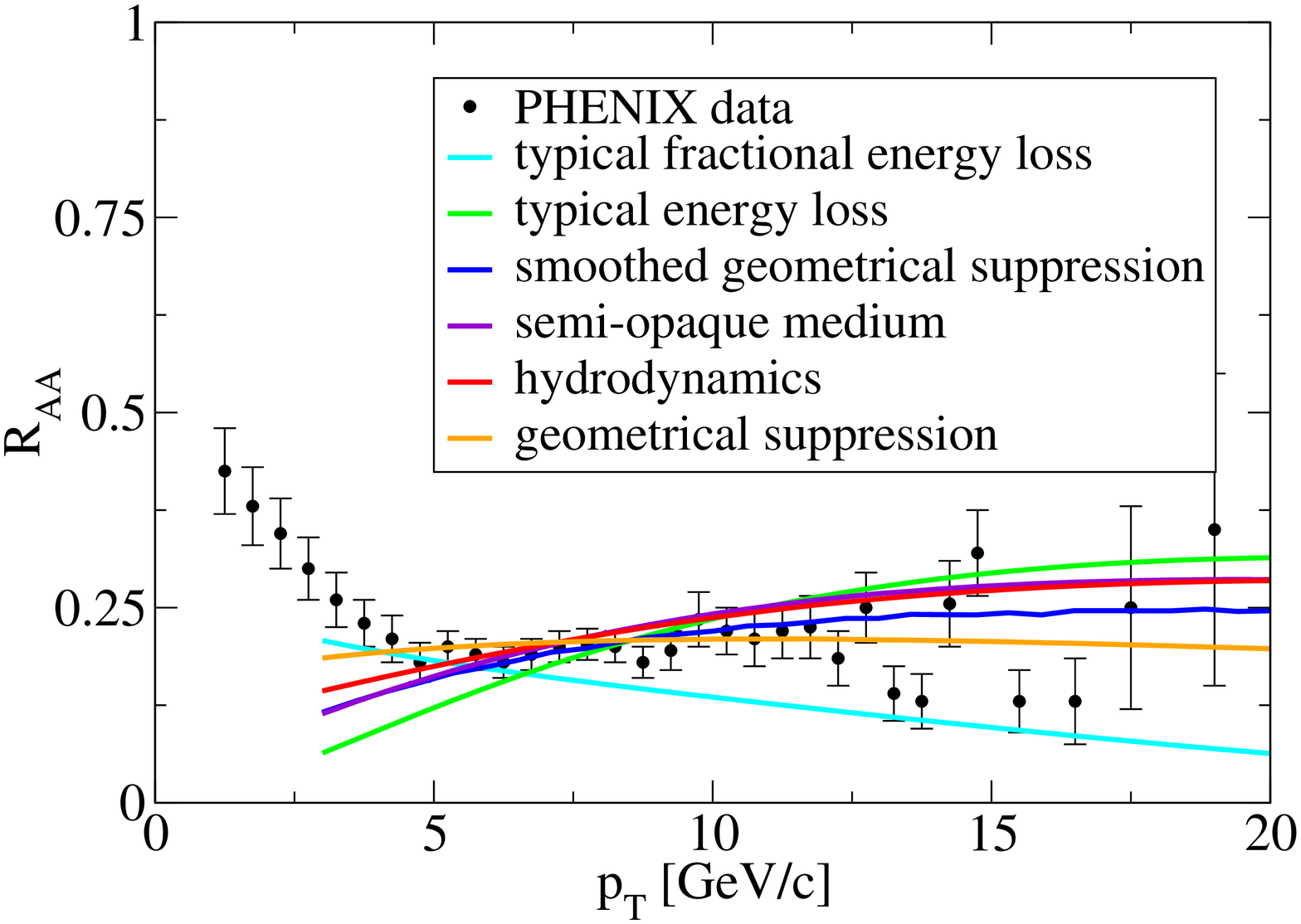, width=8cm}
\caption{\label{F-Comp}Left panel: Trial energy loss probability distributions $\langle P(\Delta E, E)\rangle_{T_{AA}}$ and the result of a full hydrodynamical simulation. Right: Nuclear suppression factor $R_{AA}$ for different toy models and the hydrodynamical simulation compared with PHENIX data \cite{PHENIX_R_AA}.}
\end{figure*}

Several trial energy loss probabilities are shown in Fig.~\ref{F-Comp}, left panel. The different scenarios are described in greater detail in \cite{Gamma-Hadron}. In each, one parameter has been tuned to result in the best possible description of $R_{AA}$. In addition, we also test a constant fractional energy loss in which every parton loses 30\% of its initial energy.

The resulting $R_{AA}$ after evaluating Eqs.~(\ref{E-med},\ref{E-R_AA}) is shown in Fig.~\ref{F-Comp}, right panel. Considering that all scenarios {\it must} underpredict the data below 5--6 GeV momentum as the model does not include hadronization mechanisms other than fragmentation which is known not to describe the data in this region, it is apparent that present data cannot distinguish well between the different scenarios. Given an assumption about the functional forum of the energy loss probability, the data can only constrain a scale. An exception to this is the fractional energy loss which seems to display the wrong qualitative trend. One may conclude that the explicit energy dependence of the energy loss probability distribution is therefore weak.

In detail, there are of course difference between the scenarios, reflecting the presence or absence of an explicit scale for the average energy loss and hence the implicit energy dependence of $\langle P_f(\Delta E, E)\rangle_{T_{AA}}$. The typical energy loss shows the strongest variation as a function of $p_T$ as the definition of shift and absorption changes across different $p_T$ whereas the geometrical energy loss in which partons are absorbed based on their region of origin is scale-blind and shows no variation.

\section{$\gamma$-hadron correlations}

It is difficult to exploit the scale sensitivity as a tool to distinguish between different $\langle P_f(\Delta E, E)\rangle_{T_{AA}}$ from $R_{AA}$ as Eq.~(\ref{E-med}) contains a convolution over all possible initial parton energies. This averaging clearly removes sensitivity. However, by using $\gamma$-hadron correlation measurements it is possible to create a monochromatic source of quarks propagating through the medium \cite{Gamma-Hadron,XNPhotons1,XNPhotons}.

The momentum distribution of hadrons observed back to back with a $\gamma$ (neglecting intrinsic $k_T$ smearing) can be found from

\begin{equation}
\begin{split}
\label{E-Analysis}
p_T^\gamma\frac{dN}{dp_T^h} =& \int_{z_{min}}^1 \negthickspace  \negthickspace \negthickspace \negthickspace
dz D^{vac}_{q \rightarrow h}(z,Q^2) \int_0^1  \negthickspace
df \langle P_q(f, p_T^\gamma)\rangle  \delta\left(zf - \frac{p_T^h}{p_T^\gamma}\right) \\
& = \int_0^1 df D^{vac}_{q \rightarrow h}(p_T^h/(f \cdot p_T^\gamma), Q^2) \langle P_q(f, p_T^\gamma)\rangle.\\
\end{split}
\end{equation}

\begin{figure}[htb]
\epsfig{file=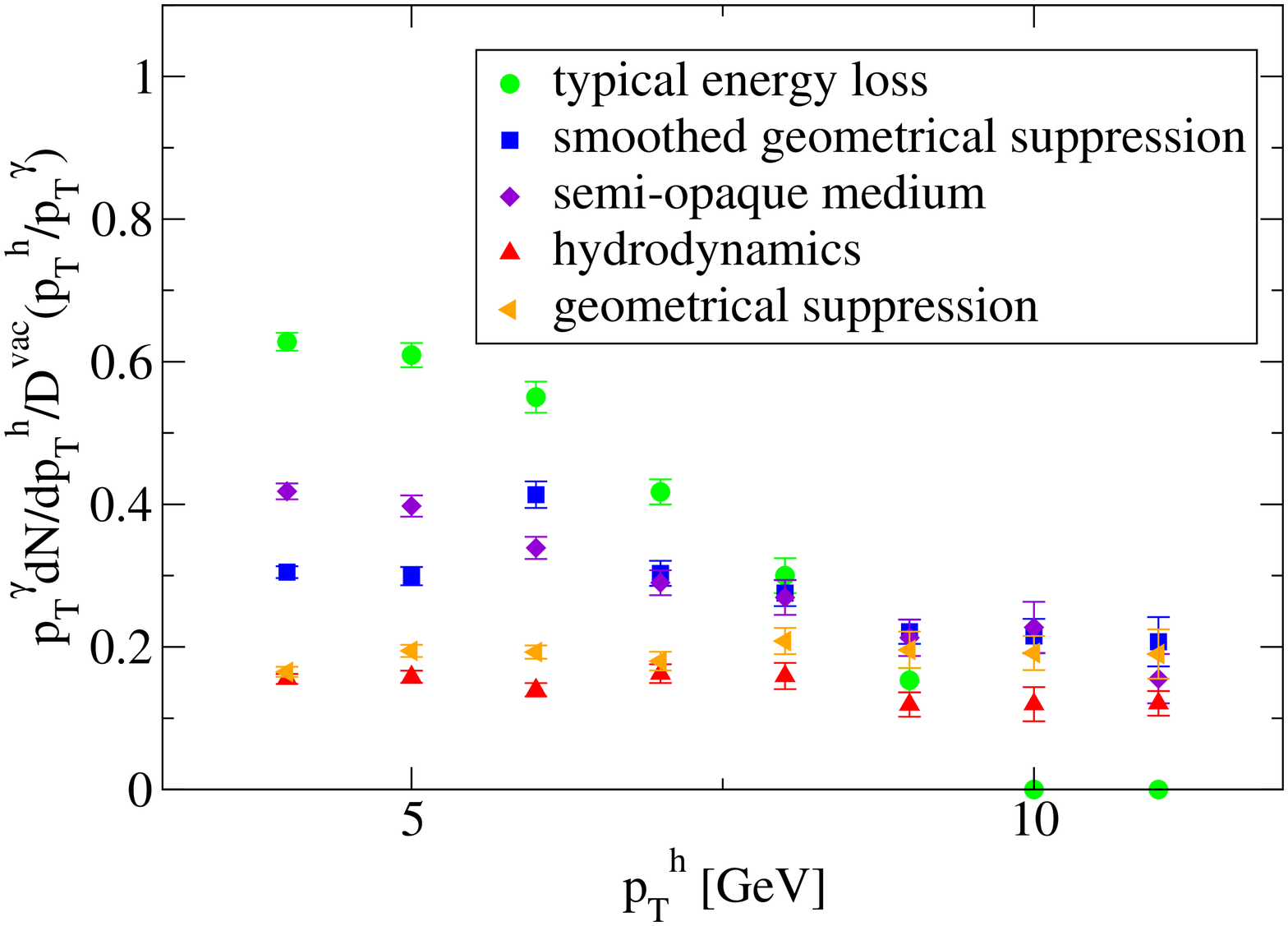, width=8cm}
\caption{\label{F-Photon}Momentum spectrum of hard hadrons correlated back-to-back with a photon trigger, normalized to the vacuum expectation.}
\end{figure}

This expression is evaluated with the trial probability distributions shown in Fig.~\ref{F-Comp} and the result is shown in Fig.~\ref{F-Photon}. As clearly seen, in such a measurement the different functional forms of the energy loss distributions can be more easily distinguished. While this technique allows in principle a model-independent measurement of $\langle P_q(\Delta E, E)\rangle_{T_{AA}}$, it is experimentally challenging and the result still does not disentangle the physics of energy loss with the geometry of density distribution. Let us therefore turn to a specific model of parton interaction with the medium and discuss what can be learned about the medium under this assumption.

\section{Radiative energy loss}

Our calculation follows the BDMPS formalism for radiative energy loss 
\cite{BDMPS} using quenching weights as introduced by
Salgado and Wiedemann \cite{QuenchingWeights}. Detailed information about the implementation is given in \cite{Correlations}, here we just present an overview.

If we call the angle between outgoing parton and the reaction plane $\phi$, 
the path of a given parton through the medium $\xi(\tau)$ is specified 
by $({\bf r_0}, \phi)$ and we compute the energy loss 
probability $P_f(\Delta E)_{path}$ for this path by 
evaluating the line integrals
\begin{equation}
\label{E-omega}
\omega_c({\bf r_0}, \phi) = \int_0^\infty \negthickspace d \xi \xi \hat{q}(\xi) \quad  \text{and} \quad \langle\hat{q}L\rangle ({\bf r_0}, \phi) = \int_0^\infty \negthickspace d \xi \hat{q}(\xi)
\end{equation}
along the path where we assume the relation
\begin{equation}
\label{E-qhat}
\hat{q}(\xi) = K \cdot 2 \cdot \epsilon^{3/4}(\xi) (\cosh \rho - \sinh \rho \cos\alpha) 
\end{equation}
between the local transport coefficient $\hat{q}(\xi)$ (specifying 
the quenching power of the medium) and energy density $\epsilon$ and the local flow velocity $\rho$ where $\alpha$ is the angle between parton propagation and flow direction.
Here, $\omega_c$ is the characteristic gluon frequency, setting the scale of the energy loss probability distribution and $\langle \hat{q} L\rangle$ is a measure of the path-length, weighted by the local quenching power.
We view 
the parameter $K$ as a tool to account for the uncertainty in the selection of $\alpha_s$ and possible non-perturbative effects increasing the quenching power of the medium (see discussion in \cite{Correlations}).

Using the numerical results of \cite{QuenchingWeights}, we obtain $P_f(\Delta E)_{path}$ 
for $\omega_c$ and $R=2\omega_c^2/\langle\hat{q}L\rangle$ as a function of jet production vertex and the angle $\phi$ (here $R$ is a dimensionless quantity needed as input for the energy loss probability distributions as defined in \cite{QuenchingWeights}).
The energy loss probability  $P_f(\Delta E)_{path}$ is derived in the limit of infinite parton energy  \cite{QuenchingWeights}. In order to account for the finite energy of the partons we truncate $P_f(\Delta E)$ 
at $\Delta E = E_{\rm jet}$ and add $\delta(\Delta-E_{\rm jet}) \int^\infty_{E_{\rm jet}} d\epsilon P(\epsilon)$. This procedure is known as non-reweighting \cite{EskolaUrs}.
We point out that the alternative concept of reweighting
to our understanding systematically overestimates $P_f(\Delta E)$ for $\Delta E \ll E_{\rm jet}$ and should be disregarded. In fact for a dense medium increasing $\hat{q}$ and employing reweighting leads to an increased escape probability whereas increasing $\hat{q}$ and non-reweighting leads to the expected decrease in escape probability, see also \cite{Blackness}.

The further strategy is then to decide on a particular model for the dynamical evolution of the medium and to adjust $K$ such that $R_{AA}$ for central collisions is reproduced (the values of $K$ needed vary about 50\% within the class of models describing bulk data, see \cite{Hydro3d}). We then look at more differential observables to study where information beyond the basic scale $K$ can be found.

\section{$R_{AA}$ for identified baryons}

Based on the observation that the older KKP fragmentation function \cite{KKP} underestimates proton production in p-p collisions measured by the STAR collaboration \cite{STAR-pp-pp} while the AKK set \cite{AKK} describes the data better, it has been suggested that the production of protons and antiprotons in hard processes is gluon-dominated. This gives rise to the expectation that proton suppression should be stronger than pion suppression, as the gluon is more strongly interacting with the medium due to its different color charge.

Since the number of produced quarks and gluons at given $p_T$ is fixed from the pQCD calculation, the only change of the model going from pion to proton production is the fragmentation function, i.e. there is no further parameter adjustment. Protonic $R_{AA}$ is therefore a non-trivial prediction of the model.

\begin{figure}
\epsfig{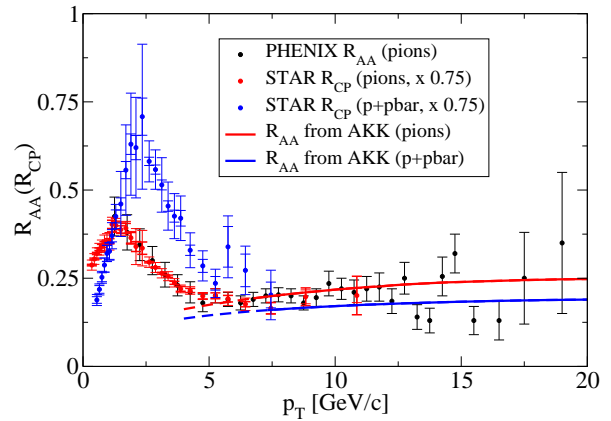}
\caption{\label{F-1}$R_{AA}$ for pions as measured by PHENIX \cite{PHENIX_R_AA}, scaled $R_{CP}$ for pions measured by STAR \cite{STAR-data} (in good agreement with the PHENIX $R_{AA}$), scaled $R_{CP}$ for protons measured by STAR \cite{STAR-data} and calculated $R_{AA}$ for pions and protons, dashed lines indicate the regime where recombination is expected to become an important effect \cite{Reco}.}
\end{figure} 

In Fig.~\ref{F-1} we show the model calculation of $R_{AA}$ with the energy loss implemented as described above for a 3d hydrodynamical model for the soft matter evolution \cite{Hydro-Chiho} for both pion and proton production and compare with PHENIX and STAR data \cite{R_AA_pp}.
Since our framework, making use of thermodynamical quantities to describe energy loss induced by the soft medium is not suited to describe $60-80$\% peripheral collisions where a dilute, predominantly hadronic, medium is expected to exist, we calculate $R_{AA}$ and scale the experimental data. We show that the scaled pionic $R_{CP}$ as obtained by STAR \cite{STAR-data} is in perfect agreement with the pionic $R_{AA}$ as obtained by PHENIX \cite{PHENIX_R_AA}. This gives us confidence that the comparison of protonic $R_{AA}$ with scaled $R_{CP}$ is not grossly unreasonable.
As apparent from the results, the measured similarity of the pionic and protonic suppression is expected from our model calculations, the difference between pion and proton $R_{AA}$ is less than the error on the pion $R_{AA}$ over a large kinematic range. We stress that we consider the calculation to be an upper limit for the expected difference as the use of the AKK set of fragmentation functions overpredicts proton production in p-p collisions \cite{STAR-pp-pp}. If this baseline is adjusted by hand, the difference between pion and proton $R_{AA}$ is reduced even further by 50\% (not shown).

\section{$R_{AA}$ vs. reaction plane}

One can also go to a more differential probe by determining the reaction plane in non-central collisions and measuring the suppression pattern of hadrons as a function of the angle between hadron and reaction plane. Since the initial overlap zone of the colliding nuclei is almond-shaped, hadron emission can be in-plane (i.e. probing the short distance of the almond) or out-of-plane (along the long axis of the almond).

In order to calculate this quantity, the angular averaging in Eq.~(\ref{E-Pav}) has to be removed, in this way, the $R_{AA}$ acquires an explicit dependence on the angle $\phi$ \cite{Hydro3d}.

\begin{figure}
\includegraphics[width=0.9\linewidth]{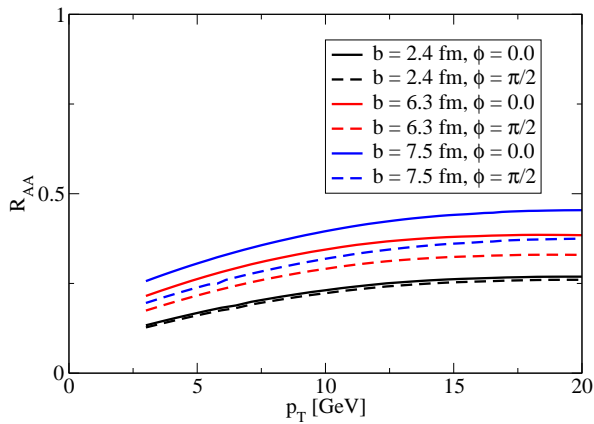}
\caption{$p_T$ dependence of $R_{AA}$ in plane (solid) and out of plane (dashed) emission at different values of impact parameter ${\bf b}$.
}
\label{fig3}
\end{figure}

In Fig.~\ref{fig3} we show the resulting $R_{AA}$ for emission in plane and out of plane for three different impact parameters for the medium specified by the 3d hydrodynamics \cite{Hydro3d}. The observed trend is as expected --- the overall quenching power decreases with increased impact parameter as the produced medium becomes less and less dense. Furthermore, an asymmetry between in-plane and out-of-plane emission develops with increased impact parameter, reflecting the increased spatial asymmetry of the initial density distribution.

\begin{figure}
\includegraphics[width=0.9\linewidth]{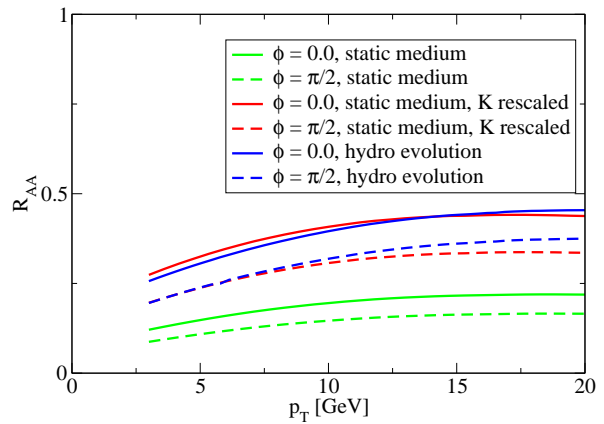}
\caption{$R_{AA}$ for in plane and out of plane emission as a function of $p_T$ at {\bf b} = 7.5 fm, assuming a static medium given by the hydrodynamical initial state, a static medium with readjusted quenching power given by the hydrodynamical initial state and the full hydrodynamical evolution. 
}
\label{fig5}
\end{figure}

However, it is well known that the pressure of the initial state creates elliptic flow which maps the initial anisotropy in position space into a corresponding anisotropy in momentum space which removes the spatial asymmetry over time. The hydrodynamical model for the soft medium describes this feature of bulk dynamics well \cite{Hydro-Chiho}. Thus, one may ask if the fact that the asymmetry is removed over time is actually reflected in the observable $R_{AA}$. In Fig.~\ref{fig5} we show a comparison of the full hydrodynamical evolution and a scenario in which we keep the initial state fixed (in this case, one has to rescale the parameter $K$ downwards to get the same description of $R_{AA}$ for central collisions as the medium does not expand and the density remains high). If one makes the comparison, one finds that elliptic flow tends to decrease the split between in-plane and out-of-plane emission as expected. In addition, there is a small distortion in shape.

A precise observation of $R_{AA}$ vs. reaction plane at sufficiently high $p_T$, taken together with the requirement that the hydrodynamics must describe the measured elliptic flow, could therefore constrain the amount of elliptic flow which is built up early on (as jet quenching is dominated by the partonic evolution phase) vs. the amount of elliptic flow built up late during the hadronic evolution.

\section{Dihadron correlations}

Instead of studying the momentum spectrum of hadrons back-to-back with a $\gamma$, one can also investigate the spectrum of associated hadrons back to back with another hadron. However, such a process shows pronounced differences.

Experimentally, a hard hadron in a given momentum range is used as the trigger and correlations are obtained with respect to this hadron. However, due to the hadronization process, it is impossible to conclude from the trigger momentum to the underlying parton momentum, rather the momentum smearing induced by the fragmentation function has to be accepted.

But second and more important, since the trigger side (near side) parton is, unlike a $\gamma$, strongly interacting, it undergoes energy loss dependent on the amount of matter it traverses. Thus, the probability density of finding vertices in the transverse plane  which lead to a triggered event is biased towards the surface of the medium. This in turn implies that when calculating the distribution of partons back-to-back with the trigger (away side),  $P_f(\Delta E)_{path}$ is not averaged using Eq.~(\ref{E-Pav}) but rather over a conditional probability distribution which becomes evident only after one has evaluated the trigger condition. Since this conditional probability density is nontrivial, it contains additional information on the medium density distribution.

We use a Monte-Carlo implementation to simulate the problem.
First, we sample the distribution of partons emerging from a  
hard vertex determined by Eq.~(\ref{E-2Partons}). This yields the parton type (quark or gluon) as  
well as the transverse momentum. We define randomly one of the partons as 'near side' and propagate  
it to the surface of the medium. Along the path, we determine $\omega_c$ and  
$\langle\hat{q}L\rangle$ by evaluating Eq.~(\ref{E-omega}). The resulting values again serve  
as input for the probability distribution of energy loss $P(\Delta E)_{path}$ as determined in  
\cite{QuenchingWeights}.

We determine the actual energy loss of the near side parton by sampling $P(\Delta E)_{path}$.
To find the energy of the leading hadron, we need the probability $P_{f \rightarrow h}(z, \mu)$ to 
find a leading hadron with momentum fraction $z$ at scale $\mu$. Strictly speaking, this is not the fragmentation function
as the fragmentation function yields the full single hadron distribution, not only the leading hadron,
but since the trigger condition enforces on average large values of $z$, the two are virtually identical
and we use the (normalized) fragmentation function $D^{vac}_{f\rightarrow h}(z, \mu)$ as a model for 
$P_{f \rightarrow h}(z, \mu)$. 
 
If the resulting hadronic $P_h = z p_f \approx z E_f$ fulfills the trigger condition we accept the  
event and proceed with the calculation of associated hadrons and the away side parton, otherwise we  
reject the event and continue the MC sampling by generating a new vertex.
 
If an event fulfilling the trigger has been created, we determine the $k_T$ smearing being added to  
the away side parton momentum. We sample a Gaussian distribution chosen such that the widening of  
the away side cone without a medium is reproduced. Since this is a number of order $1$ GeV whereas  
partons fulfilling trigger conditions have frequently in excess of $15$ GeV we note that this is a  
small correction.
 
We treat the far side parton exactly like the near side parton, i.e. we evaluate  
Eqs.~(\ref{E-omega}) along the path and find the actual energy loss from $P(\Delta E)$  
with $\omega_c, \langle\hat{q}L\rangle$ as input. If the away side parton emerges with a finite  
energy, we again use the normalized fragmentation function $D^{vac}_{f\rightarrow h}(z, \mu)$ to determine the momentum of  
the leading away side hadron. If this momentum fulfills the 
imposed trigger condition
for associated particle production, we count the event as 'punchthrough'.
 
In addition, we allow for the possibility that the fragmentation of near and away side parton  
produces more than one hard hadron. We cannot simply subsume this in the fragmentation function as done for 
single hadron distributions as we are explicitly interested in the correlation strength between near side trigger hadron
and other near side hadrons, thus we have to calculate subleading fragmentation processes separately.
The quantity we need is $P_{f \rightarrow i}(z_1, z_2, \mu)$, i.e. the conditional probability to find
a hadron $i$ from a parent parton $f$ with momentum fraction $z_2$ given that we already produced a leading hadron $h$ with
momentum fraction $z_1$. In this language, the whole jet arises from a tower of conditional probabilities for higher 
order fragmentation processes (for a different approach see \cite{Dihadron}). 
However, since we only probe the part of this tower resulting in hard hadrons, 
the treatment simplifies considerable.

Moreover, since we are predominantly interested in the quenching  
properties of the medium and not in detailed modelling of hadron distributions inside the jet, we model the next-to-leading
conditional fragmentation
probability using the measured probability distribution  
$A_i(z_F)$ of associated hadron production in d-Au collisions \cite{Dijets1, Dijets2} as a function  
of $z_T$ where $z_T$ is the fraction of the trigger hadron momentum carried by the associated  
hadron. We include a factor $\theta(E_i - E_{\rm trigger} - \Delta E - E_{\rm assoc})$ on the near  
side and $\theta(E_i - E_{\rm punch} - \Delta E - E_{\rm assoc})$ on the far side to make sure that  
energy is conserved. Note that associated production on the far side above the $p_T$ cut is only  
possible if a punchthrough occurs. We count these events as 'near side associate production' and  
'away side associated production'.
 
Thus, the yield per trigger for dihadron correlations on the near side is determined by the sum of  
all 'near side associate production' events divided by the number of events fulfilling the trigger,  
the yield per trigger on the away side is given by the sum of 'punchthrough' and 'away side  
associated production' divided by the number of events (fulfilling the near side trigger  
condition). These quantities can be directly compared to experiment.

In \cite{Correlations,Correlations1} we have investigated dihadron correlations six different scenarios for the spacetime evolution of bulk matter from a parametrized evolution model \cite{Parametrized} as well as a 2d hydrodynamical evolution model \cite{Hydro}. In the following, we will mainly discuss the 2d-hydrodynamical evolution ('hydro') where we assume that energy loss happens throughout the evolution and the same evolution ('black core') in which we confine energy loss to the partonic phase (with a corresponding rescaling of the quenching power to very large $K$). The first scenario corresponds to an extended medium with moderate quenching properties, the second to a dense core with strong quenching power surrounded by a halo from which partons can escape essentially without significant energy loss. Detailed information on all scenarios can be found in \cite{Correlations}.

\begin{figure*}[htb]
\epsfig{file=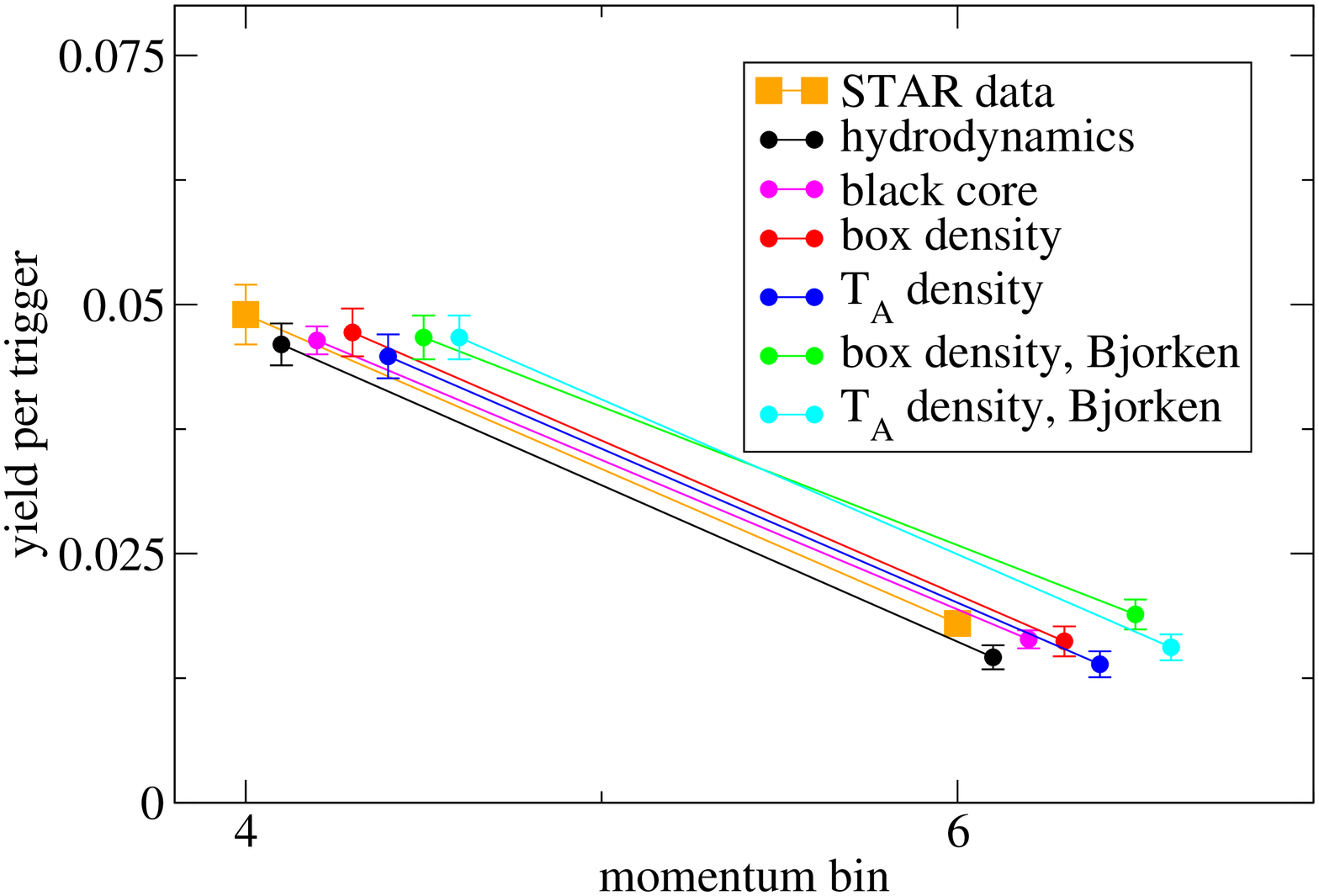, width=8cm}\epsfig{file=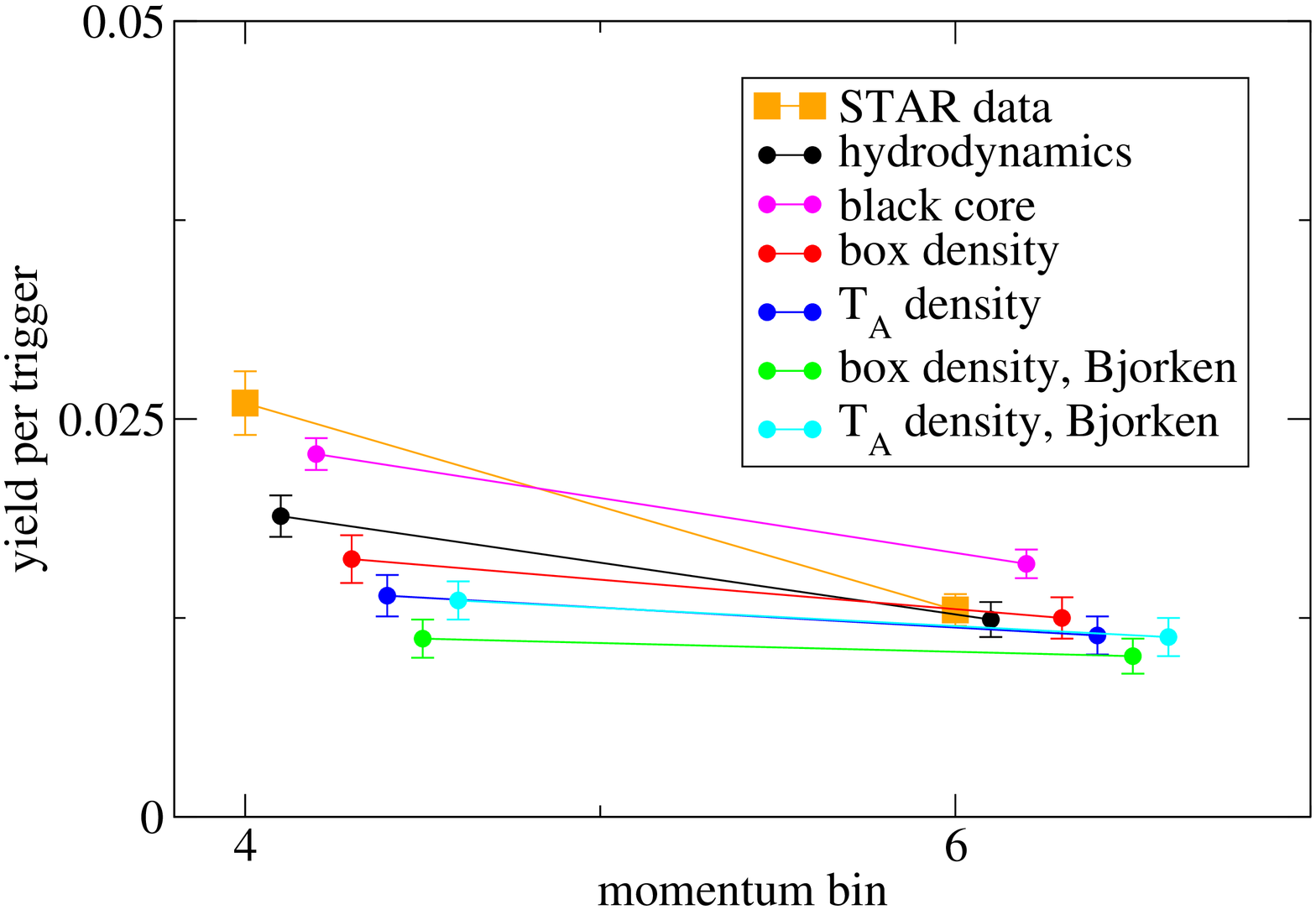, width=8cm}
\caption{\label{F-ypt8} Yield per trigger on the near side (left panel) and away side (right panel)  
of hadrons in the 4-6 GeV and 6+ GeV momentum bin associated with a trigger in the range 8 GeV $<  
p_T < $ 15 GeV for the different models of spacetime evolution as compared with the STAR data  
\cite{Dijets1,Dijets2}. The individual data points have been spread artificially along the $x$ axis  
for clarity of presentation.}
\end{figure*}

We show the calculated yield per trigger compared with the experimental result in Fig.~\ref{F-ypt8}. Within errors, the near side yield per trigger is described by all the models well. There is no 
significant disagreement among the models. 
 
The model calculations appear considerably more different if we consider the away side yield. 
Here, results for the 4-6 GeV momentum bin differ by almost a factor two.  However, none of the 
model calculations describes the data in this bin. This is in fact not at all surprising as below 
~5 GeV the inclusive single hadron transverse momentum spectra are not dominated by pQCD 
fragmentation and energy losses but, rather, by hydrodynamics possibly supplemented with 
recombination \cite{Reco,Coalescence} type phenomena.  
For this reason, the ratio $R_{AA}$ at $p_T< 5$~GeV cannot be expected to be described by pQCD 
fragmentation and energy losses, either. This is clearly unfortunate, as the model results are considerable closer to the experimental 
result in the 6+ momentum bin on the away side and hence our ability to discriminate between different models
is reduced. Since at this large transverse momenta the pQCD 
fragmentation + energy losses dominate the singe hadron spectrum we expect that the model is 
able to give a valid description of the relevant physics 
in this bin: Not only is $R_{AA}$ well described by the data, but also the contribution of 
recombination processes to the yield is expected to be small \cite{Reco}. Thus, as it stands, only 
the black core scenario can be ruled out by the data.

\begin{figure*}
\epsfig{file=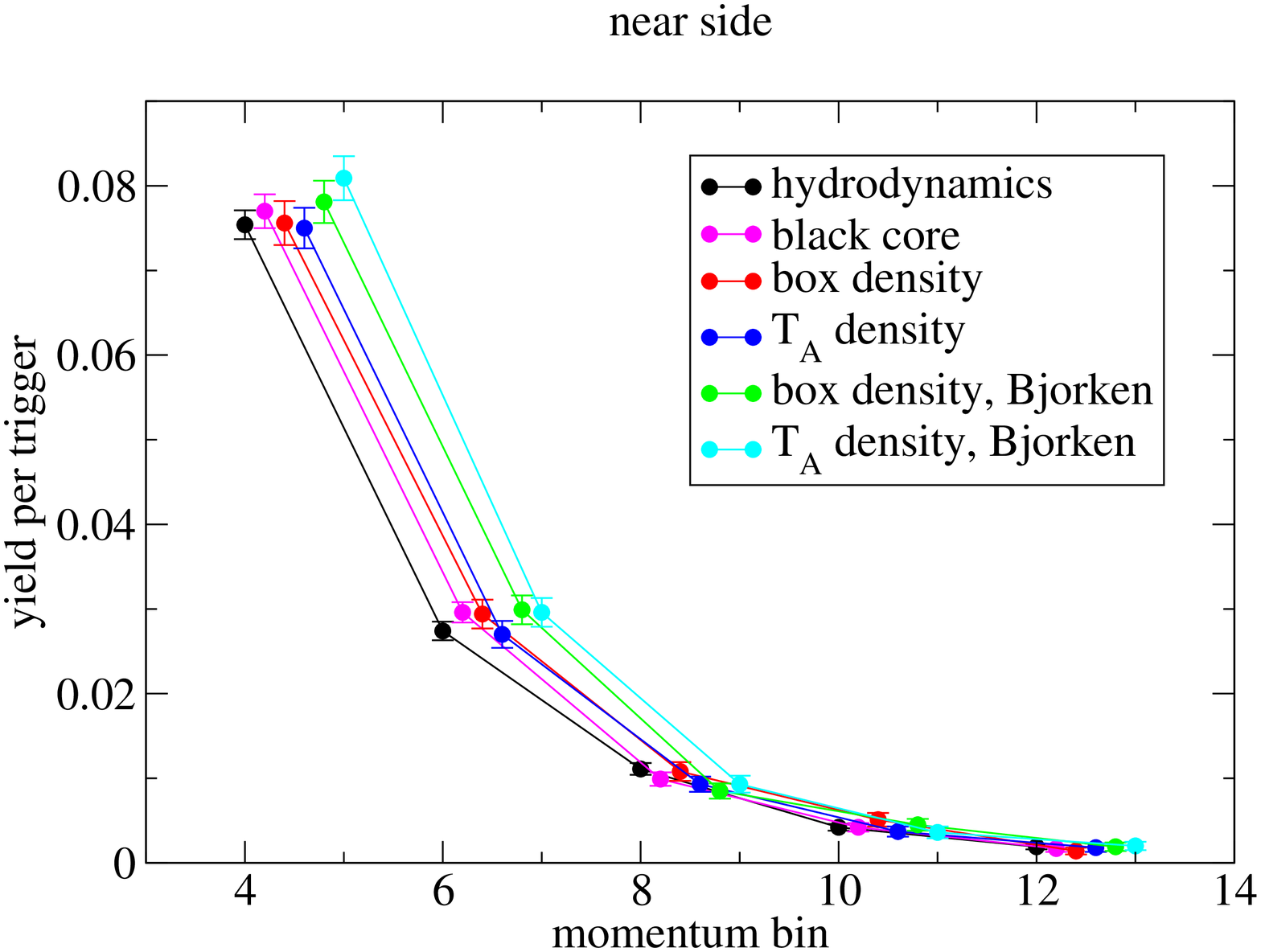, width=8cm}\epsfig{file=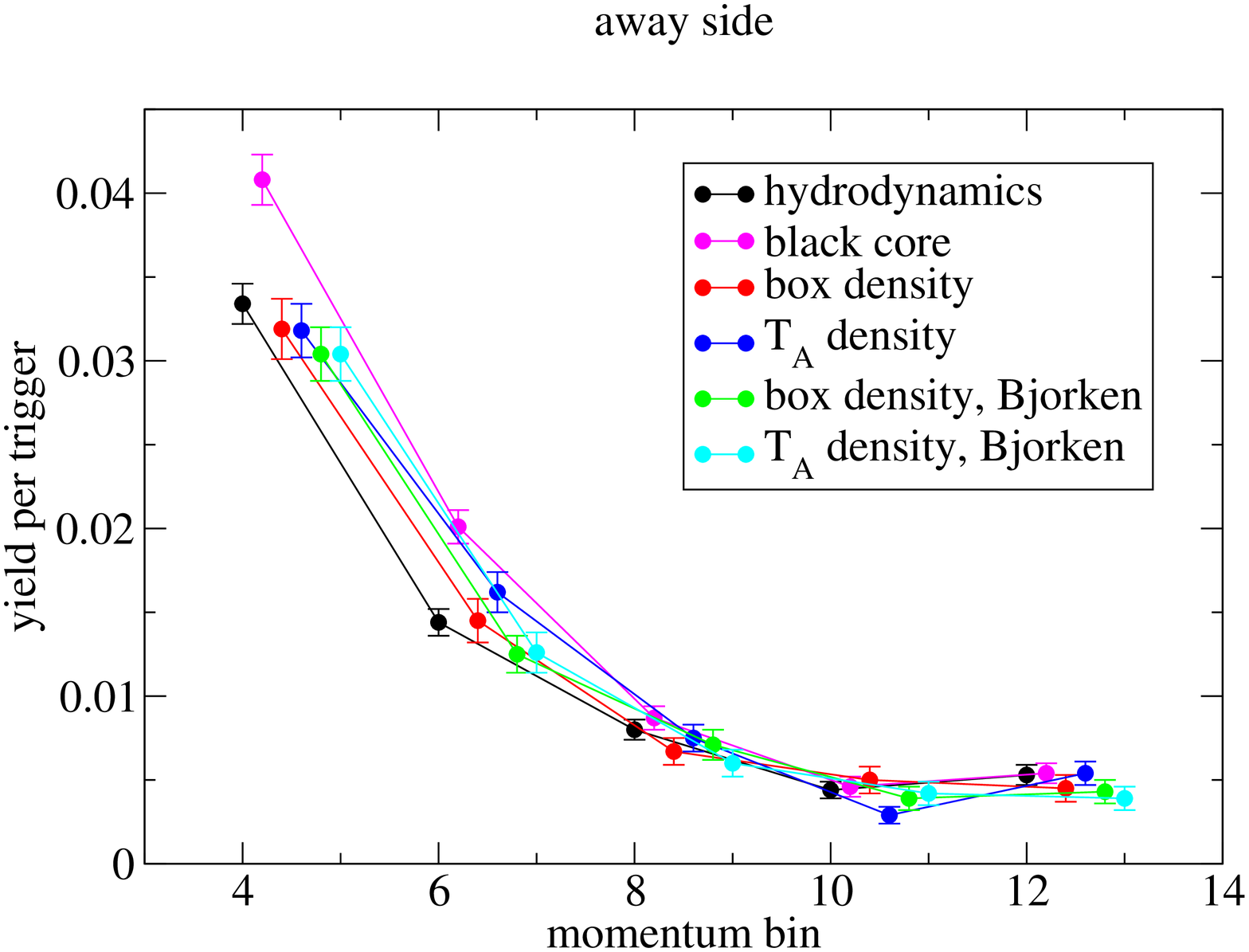, width=8cm}
\caption{\label{F-ypt12} Yield per trigger on the near side (left panel) and away side (right panel)  
of hadrons in the 4-6 GeV, 6-8 GeV, 8-10 GeV, 10-12 GeV and 12+ GeV momentum bin associated with a  
trigger in the range 12 GeV $< p_T < $ 20 GeV for the different models of spacetime evolution. The  
individual data points have been spread artificially along the $x$ axis for clarity of  
presentation. Note that the last bin extends from 12 GeV up to the $p_T$ of the trigger hadron and  
is thus considerably wider than the previous bin, explaining the upward turn of some spectra.}
\end{figure*}

Nevertheless, while there are indications that the other scenarios show sizable differences in the  
momentum spectrum of away side hadrons, the present data is not sufficient to make a distinction.  
There are in principle two ways to overcome this problem. At the price of introducing additional  
model dependence, one might include recombination processes into the simulation. In this way,  
comparison to the 4-6 GeV momentum bin would be possible. Alternatively, one can address the  
question if more leverage in $p_T$ would improve the question and hence if improved experimental  
conditions will allow tomography. We have chosen to follow the latter path. Towards this end, we  
show in Fig.~\ref{F-ypt12} a situation where the trigger momentum is increased to 12~GeV~$< p_T < $ 20  
GeV and consequently more momentum bins in the pQCD region become accessible. Clearly, the difference between 
the different evolution scenarios is somewhat magnified, although still not large.

\section{The geometry of dihadron correlations}

\begin{figure*}
\epsfig{file=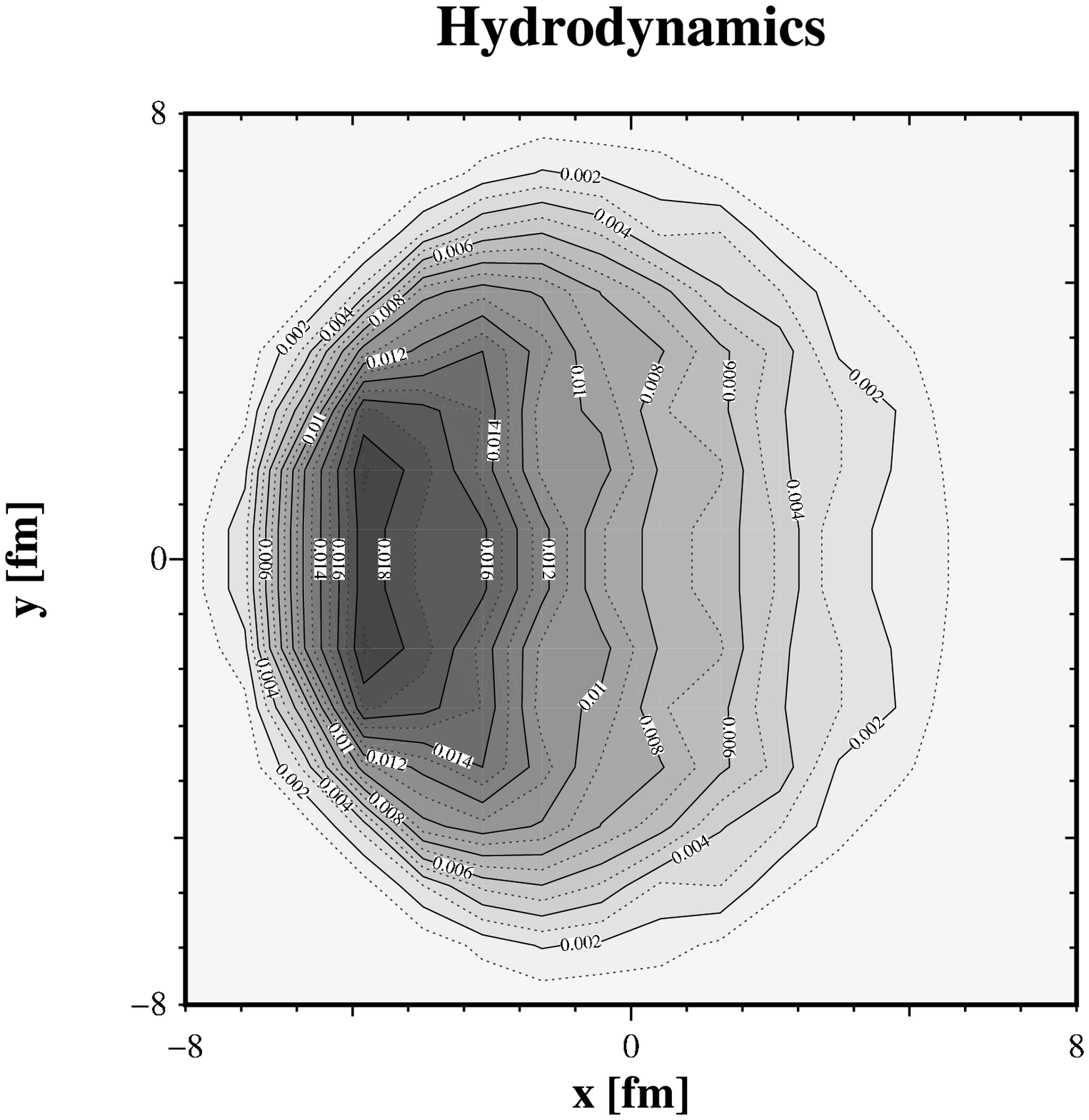, width=8cm}\epsfig{file=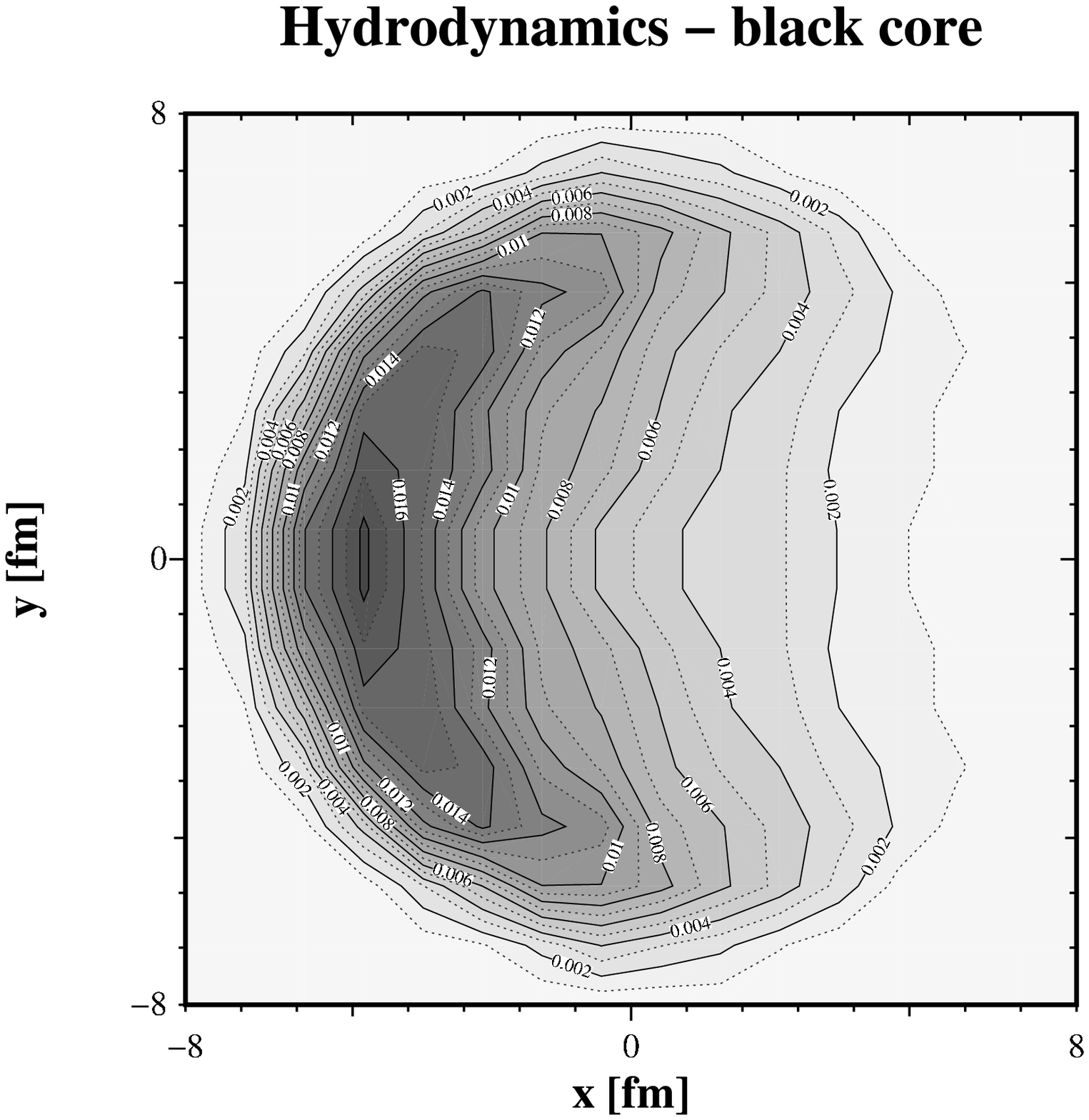, width=8cm}\\
\vspace*{-2.2cm}
\caption{\label{F-vdist} Probability density of finding a parton production vertex at $(x,y)$ 
given a triggered event with 8 GeV $< p_T <$ 15 GeV for different spacetime evolution scenarios 
(see text). In all cases the near side (triggered ) hadron propagates to the $-x$ direction, hence 
the $y- (-y)$ symmetrization. Countours are at linear intervals.}
\end{figure*}

As stated above, there is a systematical surface bias of the vertex distribution induced by the requirement to observe a trigger. For two scenarios, this distribution is shown in Fig.~\ref{F-vdist}.

It is evident that the strength of this surface bias depends on the distribution of the medium quenching power --- the surface bias of the 'black core' scenario where the central region is almost impenetrable is much stronger than in the 'hydrodynamics' scenario. Given that the in-medium pathlength of the away side parton is sampled from this distribution (rather than from Eq.~(\ref{E-PGeo})) the differences in the momentum spectrum of correlated away side hadrons in the two scenarios can be understood.

\begin{figure*}
\epsfig{file=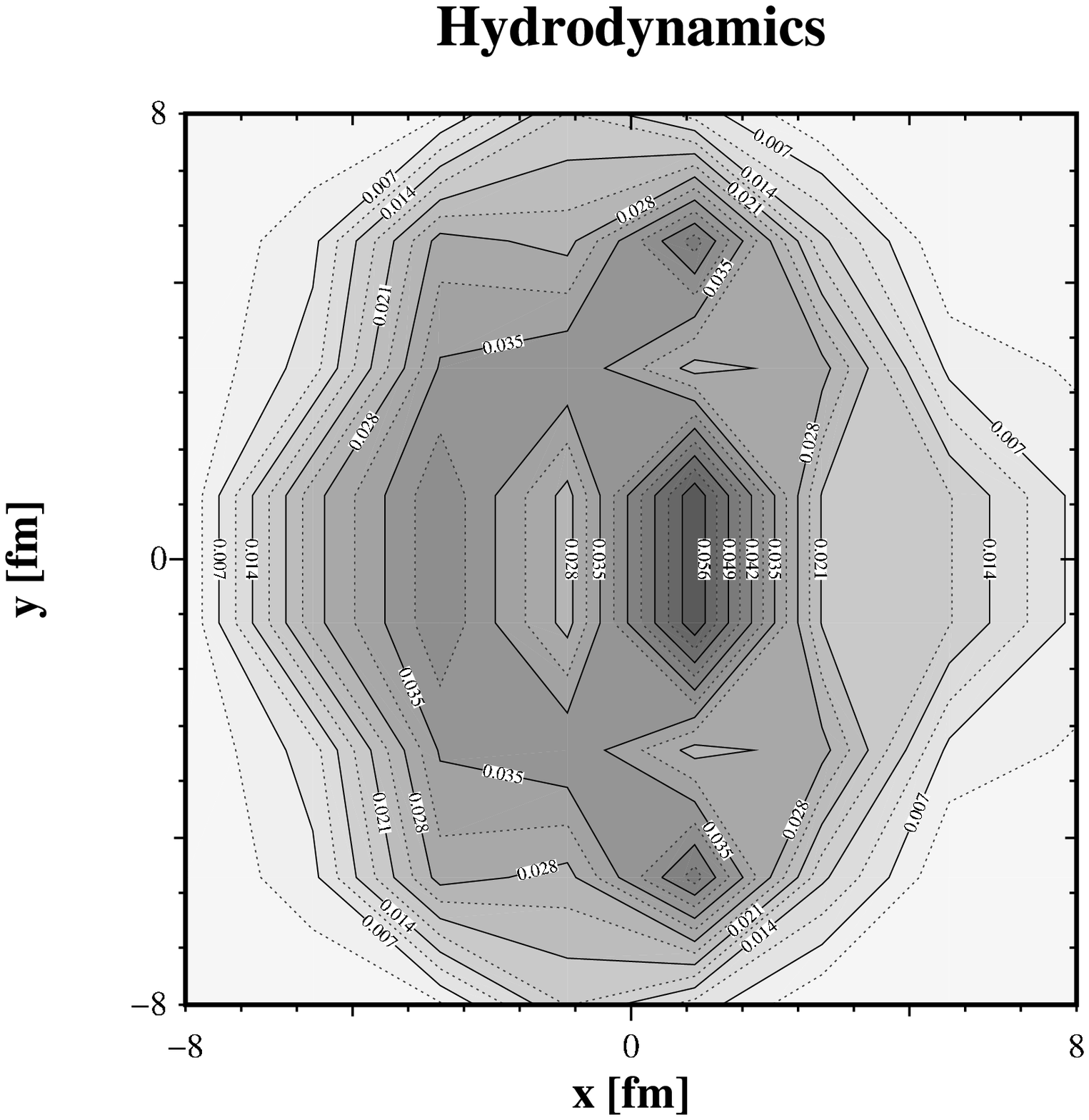, width=7.8cm}\epsfig{file=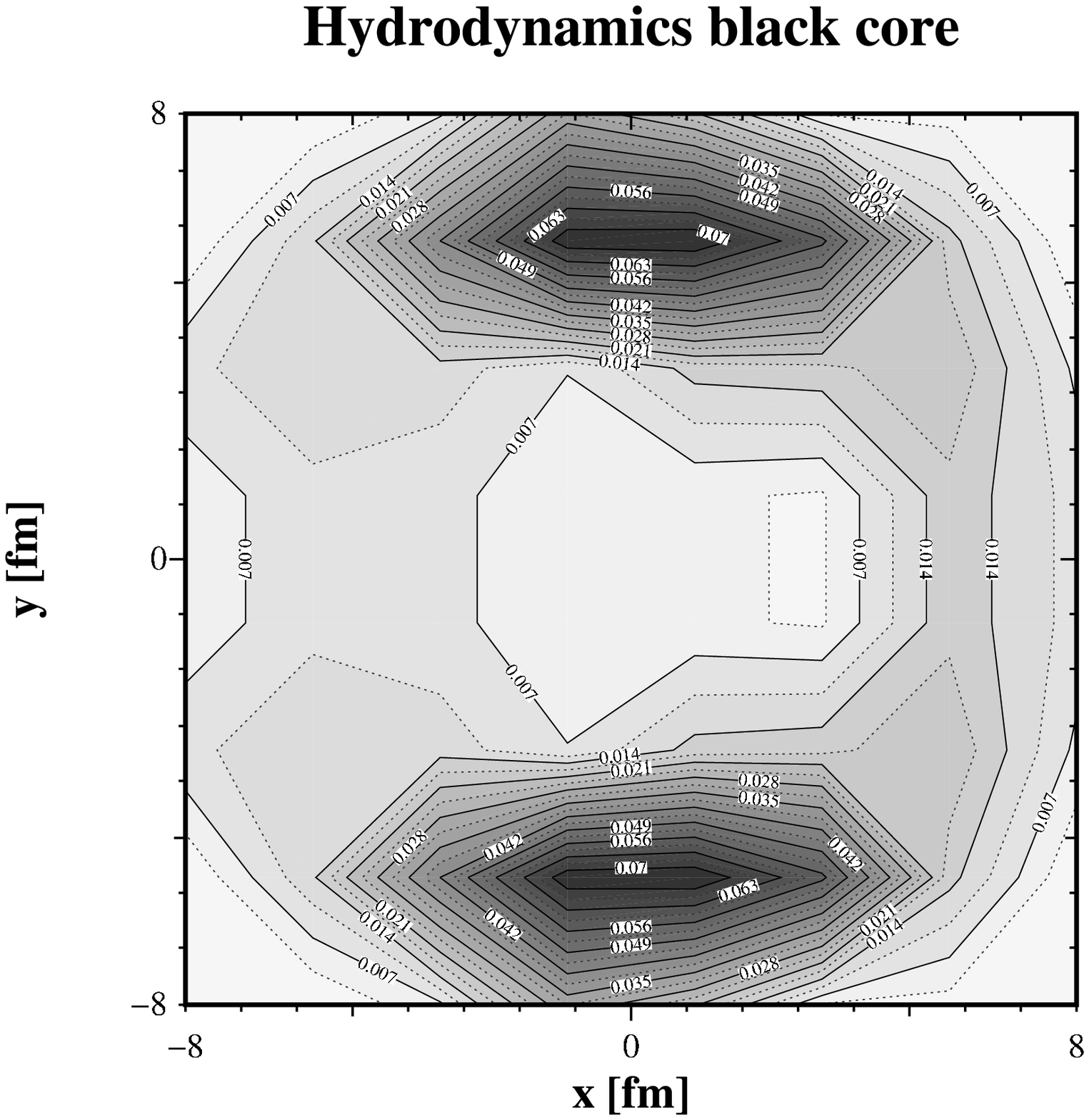, width=7.7cm}\\
\vspace*{-2.2cm}
\caption{\label{F-pdist4-6-8} Probability density for finding a vertex at $(x,y)$ leading to a 
triggered event with 8 GeV $< p_T <$ 15 GeV and in addition an away side hadron with 4 GeV $< p_T < 
6$ GeV for different spacetime evolution scenarios (see text). In all cases the near side hadron 
propagates to the $-x$ direction. Countours are at linear intervals.}
\end{figure*}

This becomes even more apparent when comparing the vertex distribution of the subclass of events in which hard hadrons were observed on both the near and away side. This is shown in Fig.~\ref{F-pdist4-6-8}. It is quite evident that the presence of a strongly quenching core leads to tangential processes in which the core is never traversed whereas in a more extended medium with moderate quenching power as in the hydrodynamics scenario the whole volume is probed.

While these figures seem compelling, it has to be kept in mind that they have been calculated under the assumption that a specific model for the interaction between hard parton and bulk medium is realized. In this sense, the calculation represents a test of particular combinations of energy loss model and medium evolution rather than a model-independent truth. From this point of view, we can note that the radiative energy loss model used here based on \cite{BDMPS,QuenchingWeights} can describe the correlation data assuming a range of hydrodynamical and hydro-inspired evolution models as long as quenching is not confined to the partonic phase alone. At present, the data are insufficient to resolve differences in detail between model evolutions.

\section{Conclusions}

We have discussed jet tomography, i.e. the idea to obtain information about the medium produced in heavy-ion collisions by studying the modifications of hard processes taking place inside the medium. We have argued that the tomographical information is contained exclusively in the energy loss probability of a parton $f$ along a specified path $P_{f}(\Delta E,E)_{path}$ if the measured hadrons are hard enough such that fragmentation can be assumed to be the dominant mechanism of hadronization and hadronization can be assumed to happen outside the medium.

Different hard observables probe different averages of this quantity. The nuclear suppression factor $R_{AA}$ for pions probes the average over all angles and the initial nuclear overlap and all parton species. By looking at proton $R_{AA}$, one may assume that predominantly gluon suppression is probed. In contrast, by using $\gamma$-hadron correlations one gets access to quark energy loss and in addition a handle on the initial energy of the parton.

In $R_{AA}$ vs. reaction plane measurements, the averaging is still performed over the full initial overlap geometry, but not over all angles which provides a systematic handle on pathlength dependence. Furthermore, low $p_T$ elliptic flow is tied to the removal of the spatial asymmetry which is reflected in the angular dependence of $R_{AA}$ and we have demonstrated that there is some sensitivity to the timescale at which this asymmetry is removed.  

Finally, back-to-back hadron correlation introduce (due to the trigger bias) a different distribution of vertices which is moved towards the surface. This implies that the away side parton is systematically biased towards long pathlengths. While difficult to control theoretically, the changed geometrical averaging implies that additional information on medium properties can be deduced from the momentum spectrum of away side hadrons.

It is quite clear that taken alone none of the measurements outlined here is able to provide a model-independent determination of the medium density. Furthermore, there is a general problem that in a situation in which spectra are steeply dropping as a function of $p_T$ observables will always be dominated by events in which no energy loss has taken place. For this reason, any effects which are sensitive to the medium density distribution are in general subleading. 

It seems that the best strategy is not to try to determine medium properties not by considering hard probes alone but to use hard probes to test models which have been used to describe soft bulk dynamics and remove possible ambiguities this way. While a single hard observable does not carry too much information, the collection of all measurable quantities sums up to to rather stringent constraints. Clearly, with both a large kinematic lever-arm for hard probes at LHC and high-statistics data from RHIC, hard probes will reach the precision needed to do tomography in the true sense of the word soon.

\begin{acknowledgments}

I would like to thank Jan Rak, Peter Jacobs, Vesa Ruuskanen and especially my collaborators J\"{o}rg Ruppert, Kari J.~Eskola, Steffen A.~Bass and Chiho Nonaka for comments, discussions and other input. This work was financially supported by the Academy of Finland, Project 115262. 

\end{acknowledgments}


\begin{thebibliography}{99}

\bibitem{Tomo1}
  M.~Gyulassy, P.~Levai and I.~Vitev,
  Phys.\ Lett.\ B {\bf 538} (2002) 282.

\bibitem{Tomo2}
  E.~Wang and X.~N.~Wang,
  Phys.\ Rev.\ Lett.\  {\bf 89} (2002) 162301.

\bibitem{Tomo3}
  C.~A.~Salgado and U.~A.~Wiedemann,
  Phys.\ Rev.\ Lett.\  {\bf 89} (2002) 092303.

\bibitem{Tomo4}
  G.~G.~Barnafoldi, P.~Levai, G.~Papp, G.~I.~Fai and M.~Gyulassy,
  Eur.\ Phys.\ J.\ C {\bf 33} (2004) S609.

\bibitem{Tomo5}
  S.~Wicks, W.~Horowitz, M.~Djordjevic and M.~Gyulassy,
  nucl-th/0512076.

\bibitem{CTEQ1}
  J.~Pumplin, D.~R.~Stump, J.~Huston, H.~L.~Lai, P.~Nadolsky and W.~K.~Tung,
  JHEP {\bf 0207}, 012 (2002).

\bibitem{CTEQ2}
  D.~Stump, J.~Huston, J.~Pumplin, W.~K.~Tung, H.~L.~Lai, S.~Kuhlmann and J.~F.~Owens,
JHEP {\bf 0310}, 046 (2003).

\bibitem{KKP}
  B.~A.~Kniehl, G.~Kramer and B.~Potter,
  Nucl.\ Phys.\ B {\bf 582}, 514 (2000).

\bibitem{AKK}
  S.~Albino, B.~A.~Kniehl and G.~Kramer,
  Nucl.\ Phys.\ B {\bf 725} (2005) 181.

\bibitem{NPDF}
  M.~Hirai, S.~Kumano and T.~H.~Nagai,
  Phys.\ Rev.\ C {\bf 70}, 044905 (2004).

\bibitem{EKS98}
  K.~J.~Eskola, V.~J.~Kolhinen and C.~A.~Salgado,
  Eur.\ Phys.\ J.\ C {\bf 9} (1999) 61.


\bibitem{EKS07}
  K.~J.~Eskola, V.~J.~Kolhinen, H.~Paukkunen and C.~A.~Salgado,
  hep-ph/0703104.


\bibitem{XN1}
  X.~f.~Guo and X.~N.~Wang,
  Phys.\ Rev.\ Lett.\  {\bf 85}, 3591 (2000).

\bibitem{XN2}
  X.~N.~Wang and X.~f.~Guo,
  Nucl.\ Phys.\  A {\bf 696}, 788 (2001).

\bibitem{Coalescence}
  R.~C.~Hwa and C.~B.~Yang,
  Phys.\ Rev.\  C {\bf 70} (2004) 024905.

\bibitem{Reco}
  R.~J.~Fries, B.~Muller, C.~Nonaka and S.~A.~Bass,
  Phys.\ Rev.\ C {\bf 68} (2003) 044902.


\bibitem{EH}
  K.~J.~Eskola and H.~Honkanen,
  Nucl.\ Phys.\  A {\bf 713} (2003) 167.

\bibitem{PHENIX_R_AA}
  M.~Shimomura  [PHENIX Collaboration],
  nucl-ex/0510023.


\bibitem{Gamma-Hadron}
  T.~Renk,
  Phys.\ Rev.\  C {\bf 74} (2006) 034906.

\bibitem{XNPhotons1}
  X.~N.~Wang, Z.~Huang and I.~Sarcevic,
  Phys.\ Rev.\ Lett.\  {\bf 77} (1996) 231.

\bibitem{XNPhotons}
  X.~N.~Wang and Z.~Huang,
  Phys.\ Rev.\ C {\bf 55} (1997) 3047.

\bibitem{BDMPS}
  R.~Baier, Y.~L.~Dokshitzer, A.~H.~Mueller, S.~Peigne and D.~Schiff,
  Nucl.\ Phys.\ B {\bf 484}, (1997) 265.

\bibitem{QuenchingWeights}
  C.~A.~Salgado and U.~A.~Wiedemann,
  Phys.\ Rev.\ D {\bf 68}, (2003) 014008.

\bibitem{Correlations}
  T.~Renk and K.~J.~Eskola, hep-ph/0610059.

\bibitem{EskolaUrs}
 K.~J.~Eskola, H.~Honkanen, C.~A.~Salgado and U.~A.~Wiedemann,
  Nucl.\ Phys.\ A {\bf 747} (2005) 511.
 
\bibitem{Blackness}  T.~Renk,
  Phys.\ Rev.\ C {\bf 74}, (2006) 024903.

\bibitem{Hydro3d}
  T.~Renk, J.~Ruppert, C.~Nonaka and S.~A.~Bass,
  Phys.\ Rev.\  C {\bf 75} (2007) 031902.

\bibitem{STAR-data}
  B.~I.~Abelev {\it et al.}  [STAR Collaboration],
  Phys.\ Rev.\ Lett.\  {\bf 97} (2006) 152301.

\bibitem{STAR-pp-pp}
  J.~Adams {\it et al.}  [STAR Collaboration],
  Phys.\ Lett.\ B {\bf 637} (2006) 161.

\bibitem{Hydro-Chiho}
  C.~Nonaka and S.~A.~Bass,
  Phys.\ Rev.\ C {\bf 75}, (2007) 014902.

\bibitem{R_AA_pp}
  T.~Renk and K.~J.~Eskola,
  hep-ph/0702096.

\bibitem{Dihadron}
  A.~Majumder, E.~Wang and X.~N.~Wang,
  nucl-th/0412061.

\bibitem{Dijets1}
  D.~Magestro  [STAR Collaboration],
  nucl-ex/0510002; talk Quark Matter 2005.
 
\bibitem{Dijets2}
  J.~Adams {\it et al.}  [STAR Collaboration],
  nucl-ex/0604018.


\bibitem{Correlations1}
  T.~Renk,
  Phys.\ Rev.\  C {\bf 74} (2006) 024903.

\bibitem{Parametrized}
  T.~Renk,
  Phys.\ Rev.\ C {\bf 70} (2004) 021903.

\bibitem{Hydro}
  K.~J.~Eskola, H.~Honkanen, H.~Niemi, P.~V.~Ruuskanen and S.~S.~Rasanen,
  Phys.\ Rev.\ C {\bf 72} (2005) 044904.

\end{thebibliography}
\end{document}